\documentclass[twocolumn,showpacs,preprintnumbers,superscriptaddress,aps]{revtex4-1}

\usepackage{graphicx,color}

\usepackage{dcolumn}

\usepackage{bm}
\usepackage{here}
\usepackage[]{lineno}
\usepackage{mathrsfs} 

\begin{document}

\preprint{}

\title{Large Anomalous Hall Conductivity Derived from \\
an $f$-Electron Collinear Antiferromagnetic Structure}

\author{Hisashi Kotegawa}
\affiliation{Department of Physics, Kobe University, Kobe, Hyogo 657-8501, Japan}

\author{Hiroto Tanaka}
\affiliation{Department of Physics, Kobe University, Kobe, Hyogo 657-8501, Japan}

\author{Yuta Takeuchi}
\affiliation{Department of Physics, Kobe University, Kobe, Hyogo 657-8501, Japan}

\author{Hideki Tou}
\affiliation{Department of Physics, Kobe University, Kobe, Hyogo 657-8501, Japan}

\author{Hitoshi Sugawara}
\affiliation{Department of Physics, Kobe University, Kobe, Hyogo 657-8501, Japan}

\author{Junichi Hayashi}
\affiliation{Muroran Institute of Technology, Muroran, Hokkaido 050-8585, Japan}

\author{Keiki Takeda}
\affiliation{Muroran Institute of Technology, Muroran, Hokkaido 050-8585, Japan}

\date{\today}

\begin{abstract}
Appropriate symmetry breaking generates an anomalous Hall (AH) effect, even in antiferromagnetic (AFM) materials. Itinerant magnets with $d$ electrons are typical examples that show a significant response. By contrast, the process by which a response emerges from $f$-electron AFM structures remains unclear. In this study, we show that an AFM material, Ce$_2$CuGe$_6$, yields a large AH conductivity (AHC) of $550$ $\Omega^{-1}$cm$^{-1}$, which exceeds the values previously reported in $d$-electron AFM materials. Observed features, including the scaling relation against electrical conductivity, suggest that this AH transport is induced cooperatively by both intrinsic and extrinsic mechanisms derived from the AFM structure.
\end{abstract}

\maketitle

The symmetry argument has clarified that an anomalous Hall effect (AHE) occurs with specific antiferromagnetic (AFM) structures, irrespective of intrinsic and extrinsic mechanisms \cite{Chen14}.
These types of AFM structures are represented by ferromagnetic (FM) point groups, which do not guarantee symmetric full compensation of spin configuration \cite{Chen14,Suzuki17,Smejkal20}.
For the intrinsic mechanism, which is described by the Berry curvatures in momentum space \cite{Chang,Sundaram,Onoda,Jungwirth,Nagaosa10}, a major factor in obtaining large responses is the off-diagonal term of the Bloch states between different bands, which are affected strongly by spin--orbit interaction.
This endows the conduction electrons with transverse anomalous velocity according to their spin directions \cite{Nagaosa10,KL}. Therefore, itinerant magnets are naively considered suitable to generate this type of response.
In fact, Fe is a typical example that exhibits a large intrinsic AH conductivity (AHC) of $>1000$ $\Omega^{-1}$cm$^{-1}$ for ferromagnets \cite{Miyasato2007}. For AFM materials, Mn$_3$$Z$ ($Z=$ Sn, Ge) and NbMnP, in which the conduction electrons mainly originate from magnetic $3d$ electrons, exhibit large AHCs of $100-450$ $\Omega^{-1}$cm$^{-1}$ by their intrinsic nature \cite{Nakatsuji2015,Kiyohara16,Nayak16,Kotegawa_NbMnP,Arai_NbMnP}.
However, symmetrical consideration, which restricts the magnetic point group, does not indicate whether the magnetism originates from itinerant electrons or localized electrons.
$4f$ electrons have strong spin--orbit coupling, which is advantageous for large Berry curvatures. However, they are generally not conductive.
Whether the spin texture of localized $4f$ electrons can offer large Berry curvatures to the conduction electrons via exchange interaction is an open question. 
Another question is whether an extrinsic mechanism, which originates in skew- and side-jump scattering, contributes to AHC in AFM structures.
In $5f$-electron ferromagnets, the extrinsic contribution has been remarkably observed together with the intrinsic contribution \cite{Siddiquee}.
The extrinsic mechanism should also be effective under an AFM structure, the symmetry of which is equivalent to ferromagnetism \cite{Chen14}.
However, the magnitude of the AHC remains unclear.
The manner in which a Hall response is generated by an $f$-electron AFM structure is intriguing. 
However, this type of example has yet to be reported.

\begin{figure}[b]
\includegraphics[width=0.8\linewidth]{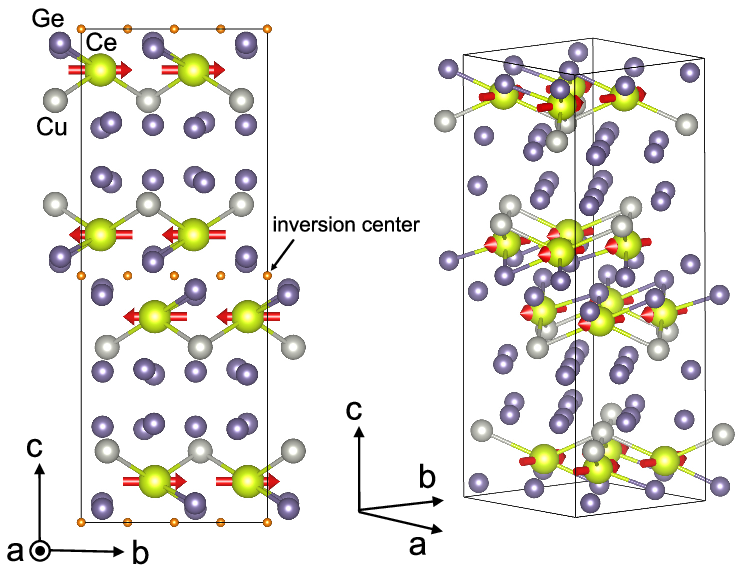}
\caption{Crystal and magnetic structure of Ce$_2$CuGe$_6$ in the orthorhombic $Cmce$ space group \cite{Qi}. The $Q=0$ magnetic structure is represented by a $B_{1g}$ irreducible representation, which corresponds to the $m'm'm$ magnetic point group. This magnetic point group allows the FM component to emerge along the $c$ axis, whereas the main components of 2 $\mu_B$/Ce are directed along the $b$ axis to form a collinear AFM structure. The orange points in the left panel indicate the inversion centers of the crystal. This AFM alignment of the Ce moments breaks the $\mathcal{PT}$ symmetry.}
\label{structure}
\end{figure}

Here, we introduce Ce$_2$CuGe$_6$, whose AFM structure is described by the magnetic point group that allows an FM state \cite{Qi}. 
Ce$_2$CuGe$_6$ crystalizes in an orthorhombic structure with the $Cmce$ (or $Cmca$) space group (No. 64).
This shows that the AFM transition at $T_{\rm N}\simeq15$ K is accompanied by a small net magnetization of $\sim10^{-2}$ $\mu_B$/Ce along the $c$ axis \cite{Sologub,Nakashima,Yaguchi,Qi}.
The magnetic structure with the propagation vector $Q=0$ is represented by the $\Gamma_3$ irreducible representation, which corresponds to the $B_{1g}$ representation or the magnetic point group $m'm'm$ \cite{Qi}.
This symmetry induces AFM components in the $ab$ plane and FM components along the $c$ axis, thereby generating the AHE even in an in-plane AFM structure.
The actual magnetic structure of Ce$_2$CuGe$_6$ is shown in Fig. 1 \cite{Qi}, which is illustrated by {\it VESTA} \cite{VESTA}.
The $b$-axis AFM components of 2 $\mu_B$/Ce are dominant and consist of the AFM stacking of FM double layers.
The Ce sites connected by the inversion symmetry, the center of which is located in the $z=0$ and $0.5$ planes, exhibit FM coupling. Accordingly, this magnetic structure preserves the space-inversion symmetry $\mathcal{P}$, but the $\mathcal{PT}$ symmetry is broken.
Here, $\mathcal{T}$ indicates time-reversal symmetry.
The absence of the $\mathcal{PT}$ symmetry is crucial for symmetry breaking to induce AHE. Note that the nonmagnetic atoms are not essential for the symmetry breaking, because the Ce atoms are located at a low-symmetric site (Wyckoff: 16g).
The large ordered moment of $2\mu_B$/Ce and the magnetic entropy of $\sim R \ln2$ at $T_{\rm N}$ indicate the localized character of the $4f$ electrons \cite{Qi,Nakashima}. 
Therefore, Ce$_2$CuGe$_6$ is a suitable material to investigate how a Hall response is derived from the localized AFM structure.
Another advantage in using this system is the large variety of similar crystal structures \cite{Sologub}.
For example, Ce$_2$PdGe$_6$ and Ce$_2$AuGe$_6$ crystalize in the same structure as that of Ce$_2$CuGe$_6$.
Ce$_2$PdGe$_6$ shows the AFM transition at $T_{\rm N}\simeq11.5$ K, where its magnetic structure has yet to be revealed. However, the same magnetic structure as that of Ce$_2$CuGe$_6$ can be expected from the similar magnetic character \cite{Yaguchi}.
The physical properties of Ce$_2$AuGe$_6$ remain unknown.

In this study, we report Hall resistivity measurements for AFM materials Ce$_2$$T$Ge$_6$ ($T=$ Cu, Pd, and Au).
We observe clear hysteresis in the field sweep of the Hall resistivities for $T=$ Cu and Pd, i.e., zero-field AHE.
The estimated AHC at the lowest temperature is $\sim550$ $\Omega^{-1}$cm$^{-1}$ for Ce$_2$CuGe$_6$ and $\sim100$ $\Omega^{-1}$cm$^{-1}$ for Ce$_2$PdGe$_6$, which are comparable to those observed in ferromagnets.
Our results show that the $f$-electron AFM structure generates a large Hall response cooperatively through both intrinsic and extrinsic mechanisms.

The platelike single crystals of Ce$_2$$T$Ge$_6$ ($T=$ Cu, Pd, and Au) are synthesized using a Bi-flux method from the ratio of Ce : $T$ : Ge : Bi = 2 : 1 : 6 : 30 \cite{Nakashima,Yaguchi}.
The single crystal of Ce$_2$AuGe$_6$ is grown for the first time.
After generating electrical contacts for gold wires using the spot weld method, we measured the electrical and Hall resistivities using a standard four-probe method.
The Hall resistivity was antisymmetrized against magnetic fields to remove the longitudinal component induced by contact misalignment. 
We measured magnetization in the temperature range of $2-300$ K and the field range up to $\pm5$ T using a Quantum Design magnetic property measurement system.

\begin{figure}[htb]
\includegraphics[width=0.85\linewidth]{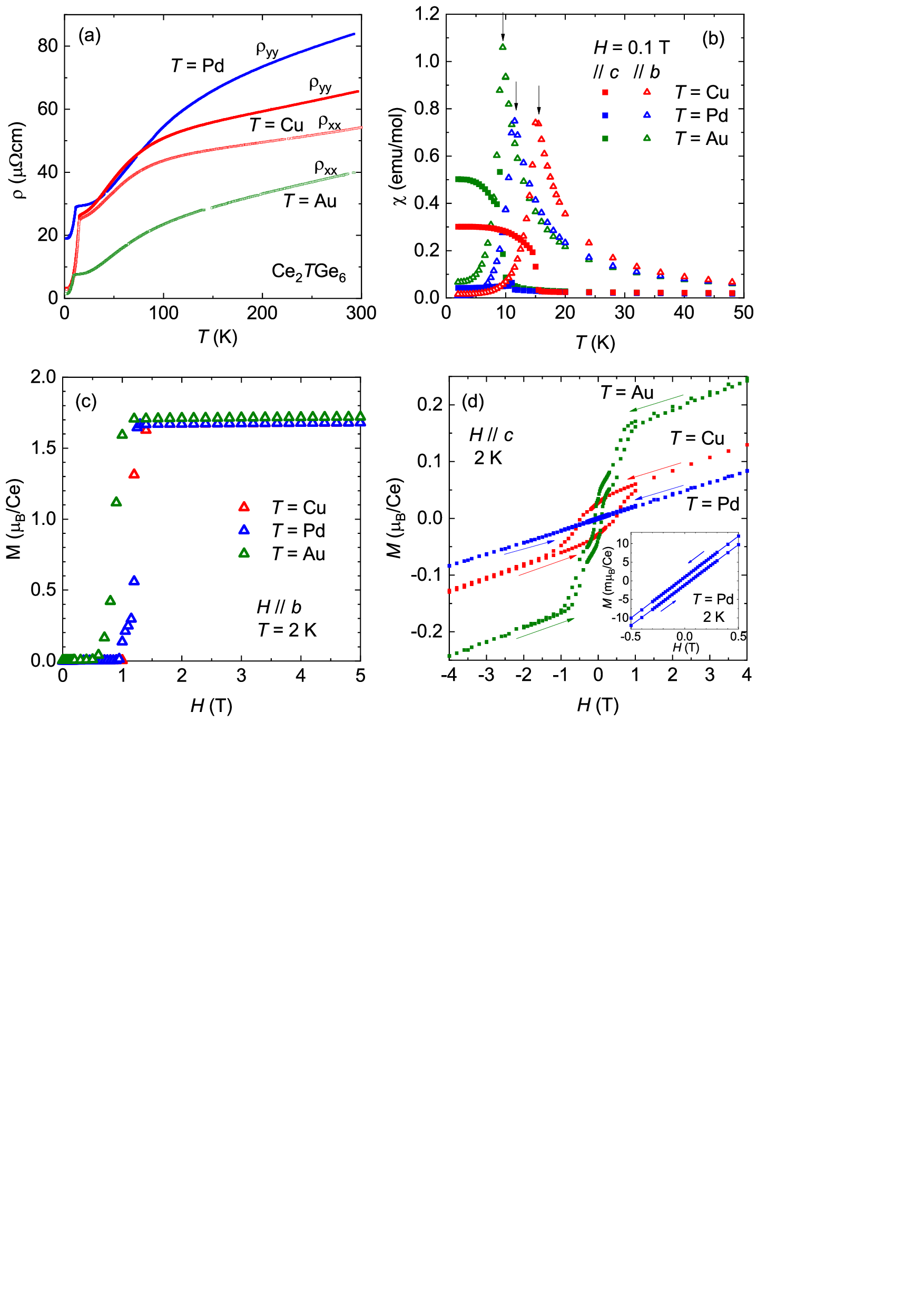}
\caption{(a) Electrical resistivities for Ce$_2$$T$Ge$_6$ ($T=$ Cu, Pd, and Au). Clear AFM transitions at $T_{\rm N}$ can be observed. The $T_{\rm N}$ was estimated to be 15.0 K for $T$= Cu, 11.3 K for $T=$ Pd, and 9.3 K for $T=$ Au. Anisotropy between $\rho_{xx}$ and $\rho_{yy}$ was weak. (b) Temperature dependence of magnetic susceptibility for three compounds. For $H \parallel b$, typical AFM-like behaviors are exhibited, whereas the FM components were induced for $H \parallel c$ for all compounds. (c,d) Magnetization curves at 2 K for $H \parallel b$ and $H \parallel c$. The metamagnetic transitions at approximately 1 T along the $b$ axis for three compounds suggests they have a similar magnetic structure. For $H \parallel c$, small spontaneous magnetizations were induced: $3\times10^{-2}$ $\mu_B$/Ce for Ce$_2$CuGe$_6$, $1\times10^{-3}$ $\mu_B$/Ce for Ce$_2$PdGe$_6$, and $4\times10^{-2}$ $\mu_B$/Ce for Ce$_2$AuGe$_6$.}
\end{figure}

Figure 2(a) shows the temperature dependences of the electrical resistivities for Ce$_2$$T$Ge$_6$ ($T=$ Cu, Pd, and Au).
They all exhibited a broad shoulder at approximately 80 K, most likely due to the crystal-field splitting of the $4f$ multiplet, which was followed by clear decreases below $T_{\rm N}$.
Anisotropy between $\rho_{xx}$ and $\rho_{yy}$, which was checked for Ce$_2$CuGe$_6$, was weak, particularly when below $T_{\rm N}$. 
Here, the $a$, $b$, and $c$ axes correspond to $x$, $y$, and $z$, respectively.
The residual resistivity of Ce$_2$CuGe$_6$ ($\rho_0=3$ $\mu\Omega$cm) was comparable to that in the previous study \cite{Nakashima}, and 
the residual resistivities of other compounds were $\rho_0=19$ $\mu\Omega$cm for $T=$ Pd and $\rho_0=1.5$ $\mu\Omega$cm for $T=$ Au.
As Fig.~2(b) shows, the magnetic susceptibility for $H \parallel b$ showed clear kinks at $T_{\rm N}$, whereas increases below $T_{\rm N}$ were observed for $H \parallel c$ \cite{Yaguchi}. 
Figures 2(c) and 2(d) show magnetization curves at 2 K for $H \parallel b$ and $H \parallel c$, respectively.
For $H \parallel b$, all the compounds showed metamagnetic transition at approximately 1 T, above which magnetization of $\sim1.7$ $\mu_B$/Ce was comparable to the ordered moment of 2 $\mu_B$/Ce for Ce$_2$CuGe$_6$ \cite{Qi}.
Under $H \parallel c$, hysteresis appeared with all compounds with small spontaneous magnetization: $3\times10^{-2}$ $\mu_B$/Ce for $T=$ Cu, $1\times10^{-3}$ $\mu_B$/Ce for $T=$ Pd, and $4\times10^{-2}$ $\mu_B$/Ce for $T=$ Au.
These small net magnetizations along the $c$ axis are allowed in the magnetic point group $m'm'm$ \cite{Qi}.
The similar magnetic properties among Ce$_2$$T$Ge$_6$ ($T=$ Cu, Pd, and Au) suggested that the magnetic structures for the three compounds similarly contained the $m'm'm$ component.
The observed spontaneous magnetizations corresponded to $3.5$ mT for Ce$_2$CuGe$_6$, 0.1 mT for Ce$_2$PdGe$_6$, and 5 mT for Ce$_2$AuGe$_6$, which were comparable to those in Mn$_3$$Z$ ($Z=$ Sn, Ge); that is, they were too small to yield large AHCs \cite{Manyala2004,Chen2021}.

\begin{figure*}[htb]
\includegraphics[width=0.85\linewidth]{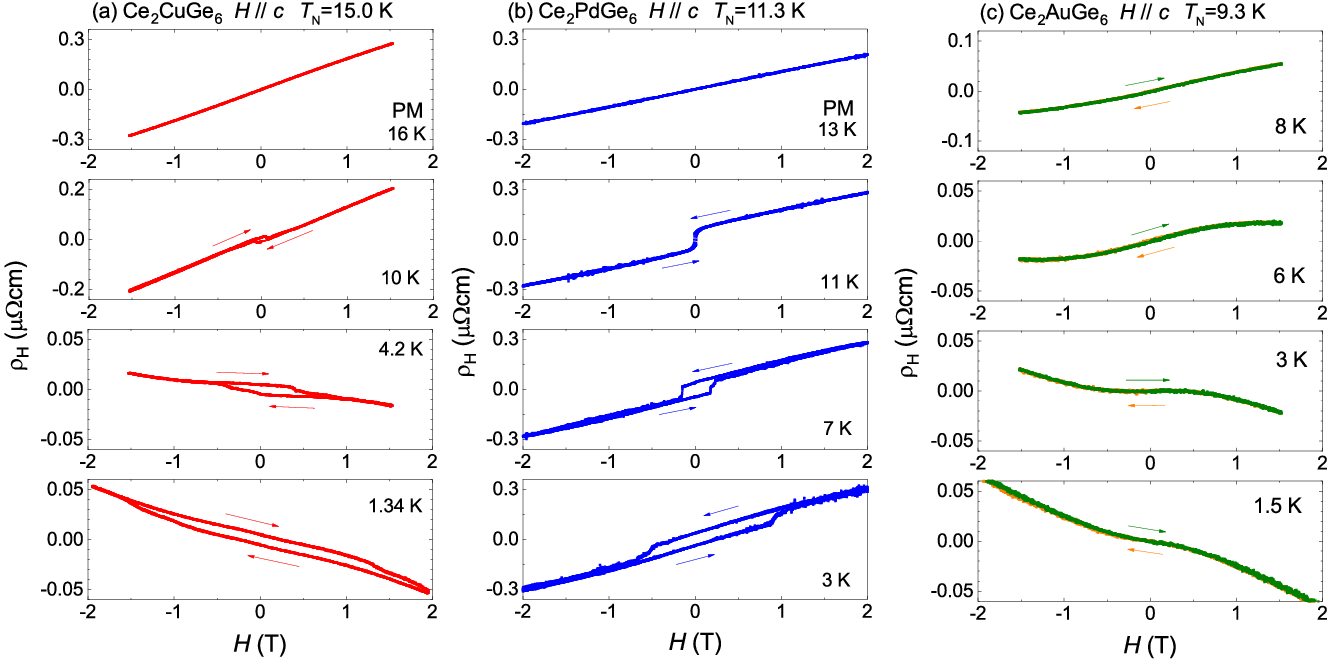}
\caption{(a)- (c) Field dependences of Hall resistivities $\rho_{\rm H}=\rho_{yx}=-\rho_{xy}$ for $T=$ Cu, Pd, and Au. In the PM state, a linear field dependence of $\rho_{\rm H}$ was observed, whereas a clear hysteresis appeared below $T_{\rm N}$ for $T=$ Cu and Pd. These zero-field AHEs show the opposite-sign values between $T=$ Cu and Pd. In $T=$ Au, the green and orange curves show the field sweep in the opposite direction. Absence of the zero-field AHE in $T=$ Au with the largest spontaneous magnetization suggested that the AHE observed in $T=$ Cu and Pd did not arise from the net magnetization.}
\end{figure*}

Figures~3(a)-(c) show the field dependences of the Hall resistivities $\rho_{\rm H}=\rho_{yx}=-\rho_{xy}$, which could be nonzero from the magnetic point group $m'm'm$.
To switch the AFM domains coupled with small net magnetization, magnetic fields were applied along the $c$ axis.
In the paramagnetic (PM) state above $T_{\rm N}$, a linear field dependence of $\rho_{\rm H}$ was observed.
Below $T_{\rm N}$, a clear hysteresis emerged for $T$= Cu and Pd, showing that a zero-field AHE was derived from their magnetic structure.
For $T=$ Au, the zero-field AHE was not confirmed within the experimental resolution.
A critical feature was observed at 10 K for $T=$ Cu, where the domain selected by the positive magnetic field generated the positive field-induced component and the negative zero-field AHE. 
This clearly demonstrated that the zero-field AHE could not be explained by the magnetization origin.

\begin{figure}[ht]
\includegraphics[width=0.9\linewidth]{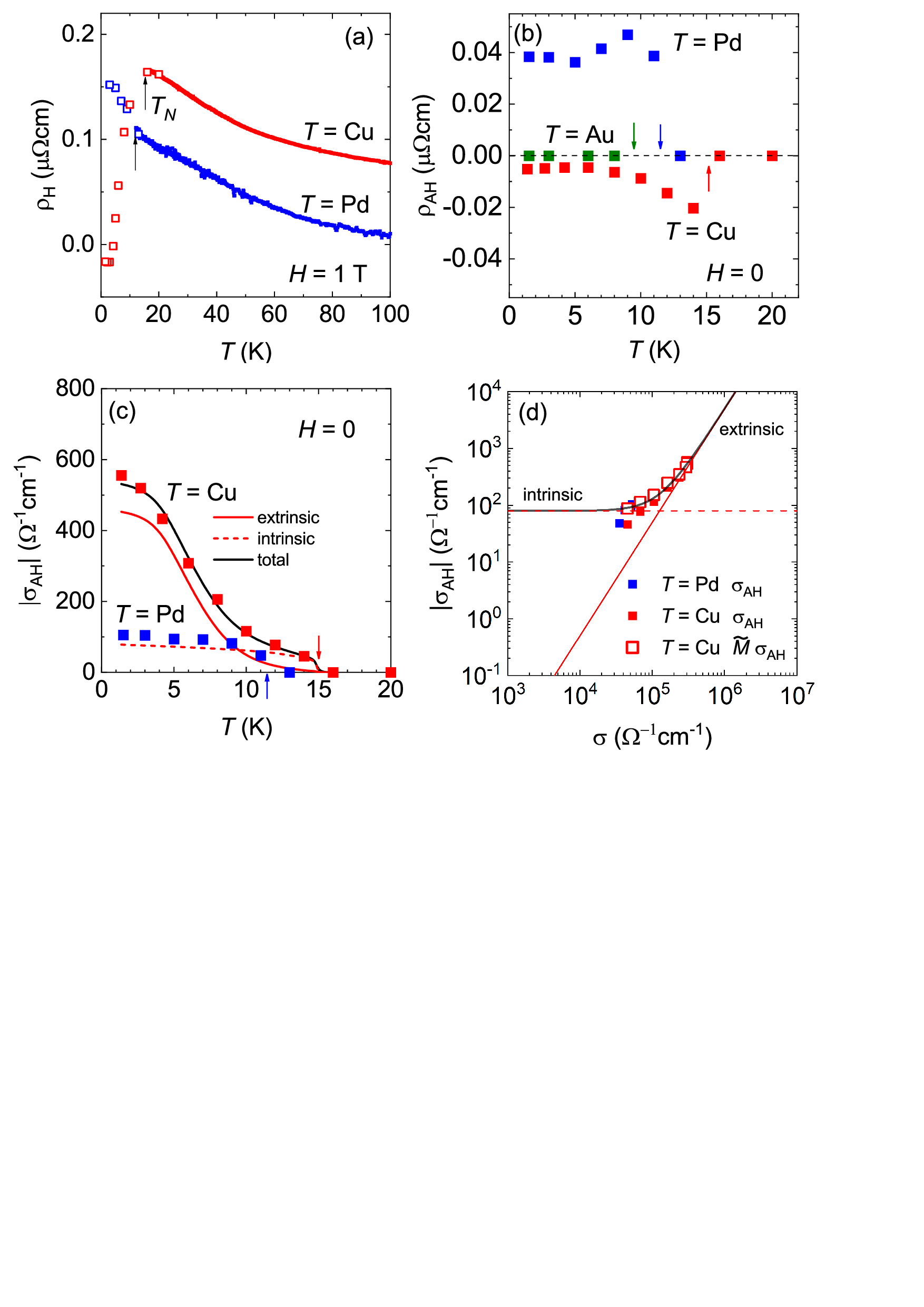}
\caption{(a) Temperature dependence of the field-induced component of $\rho_{\rm H}$, measured at $\pm 1$ T. Below $T_{\rm N}$, they were estimated from the field-sweep data by subtracting the zero-field components (open symbols). This contribution is considered to derive from incoherent skew-scattering, which has often been observed in many $f$-electron systems. (b,c) Temperature dependences of the contribution at zero-field AHE, $\rho_{\rm AH}$ and $|\sigma_{\rm AH}|$. For $T=$ Pd, $\sigma_{\rm AH}$ was estimated by assuming $\rho_{xx}\simeq\rho_{yy}$. The $|\sigma_{\rm AH}| \sim 550$ $\Omega^{-1}$cm$^{-1}$ for $T=$ Cu was higher than those of prototype $d$-electron AFM materials Mn$_3$Sn and Mn$_3$Ge.  Red dotted and solid curves indicate the contributions estimated for the intrinsic and extrinsic mechanisms for $T=$ Cu, respectively. (d) The scaling relation in $|\sigma_{\rm AH}|$ versus $\sigma$. $\widetilde{M} \sigma_{\rm AH}$  is also plotted to correct the temperature dependence of the order parameter, where $\widetilde{M}= M_{\rm AFM}(T=0)/M_{\rm AFM}(T)$. The $\widetilde{M} \sigma_{\rm AH}$ follows $\sigma_{\rm AH}^{int}+\sigma_{\rm AH}^{ext}$ (black curve), where $\sigma_{\rm AH}^{int}$ is independent of $\sigma$, and $\sigma_{\rm AH}^{ext}$ obeys $\sigma^2$. It indicates that the large $|\sigma_{\rm AH}|$ in Ce$_2$CuGe$_6$ originates in both intrinsic and extrinsic mechanisms.}
\end{figure}

Figure 4(a) shows the temperature dependence of the field-induced components in $\rho_{\rm H}$ measured at $\pm1$ T, where the zero-field components induced below $T_{\rm N}$ are not contained.
In the PM state, the positive $\rho_{\rm H}$ increased with decreasing temperature for both $T=$ Pd and Cu compounds.
These field-induced components most likely derived from incoherent skew-scattering, which has often been seen in $f$-electron systems \cite{Onuki,Fert}.
It generally follows $\rho_{\rm H} \propto \chi \rho$, where $\chi$ is the magnetic susceptibility, but the observed data did not obey this relation probably because the ordinary Hall effect also depends on temperature.
The itinerary of $f$-electrons has been known to induce a peak structure in the temperature variation of $\rho_{\rm H}$ at the onset of the coherent state \cite{Fert}.
Absence of this peak structure above $T_{\rm N}$ is qualitatively consistent with the remarkable localized character of the $f$ electrons in Ce$_2$$T$Ge$_6$.
At low temperatures, the field-induced component was significantly suppressed for $T=$ Cu, reflecting the significant decrease in electrical resistivity due to the high-quality sample.
The zero-field AHE components were estimated by $\rho_{\rm AH}=[\rho_{\rm H}(+H \rightarrow 0)-\rho_{\rm H}(-H \rightarrow 0)] /2 $ and are plotted against temperature in Fig.~4(b).
They showed nonzero values below $T_{\rm N}$ for $T=$ Cu and Pd and exhibited the maximum at slightly below $T_{\rm N}$, thus approximating a constant at low temperatures.
Absence of the anomaly in $T=$ Au with the largest net magnetization indicated that these zero-field contributions did not originate simply in the net magnetization.
The values of $\rho_{\rm AH}\simeq -5 \times 10^{-3}$ $\mu\Omega$cm for Ce$_2$CuGe$_6$ and $\rho_{\rm AH}\simeq4 \times 10^{-2}$ $\mu\Omega$cm for Ce$_2$PdGe$_6$ were much smaller than those observed in Mn$_3$Sn, Mn$_3$Ge, and low-quality NbMnP \cite{Nakatsuji2015,Kiyohara16,Nayak16,Kotegawa_NbMnP} but were comparable to that observed in high-quality NbMnP \cite{Arai_NbMnP}.
Note that a quantity determined by the Berry curvatures is not $\rho_{\rm AH}$ but $\sigma_{\rm AH}$.
The $\sigma_{\rm AH}$ is connected to $\rho_{\rm AH}$ through $\sigma_{\rm AH} = \rho_{\rm AH}/\rho^2$. 
Therefore, the intrinsic mechanism, where the dissipationless intrinsic $\sigma^{int}_{\rm AH}$ is independent of $\rho$, yields the relation of $\rho^{int}_{\rm AH} \propto \rho^2$.
Obviously, $\rho^{int}_{\rm AH}$ decreases in high-quality samples.
Figure ~4(c) shows the estimated $|\sigma_{\rm AH}|$.
The $|\sigma_{\rm AH}|$'s increased toward low temperatures, reaching maximums of $\sim550$ $\Omega^{-1}$cm$^{-1}$ for $T=$ Cu and $\sim100$ $\Omega^{-1}$cm$^{-1}$ for $T=$ Pd.
Notably, $\sigma_{\rm AH}$ for $T=$ Cu showed a steep increase below $\sim10$ K.

The value of Ce$_2$CuGe$_6$ ($\sim550$ $\Omega^{-1}$cm$^{-1}$) was higher than those of $d$-electron AFM material Mn$_3$Sn ($\sim140$ $\Omega^{-1}$cm$^{-1}$) and Mn$_3$Ge ($\sim380$ $\Omega^{-1}$cm$^{-1}$) \cite{Nakatsuji2015,Kiyohara16,Nayak16}.
To identify the origin of this large AHC through the scaling relation, the $\sigma_{\rm AH}$ versus $\sigma=1/\rho_{yy}$ is plotted in Fig.~4(d), as $\sigma_{\rm AH}^{int}$ is independent of $\sigma$ when the order parameter is fully grown \cite{Onoda2008,Nagaosa10}.
Here, we assume $\sigma_{\rm AH} \propto M_{\rm AFM}(T) \sigma^n$, where $M_{\rm AFM}(T)$ indicates the temperature variation of the order parameter in Ce$_2$CuGe$_6$ \cite{Qi}.
To identify the exponent $n$, we also plotted $\widetilde{M} \sigma_{\rm AH}$, where $\widetilde{M} = M_{\rm AFM}(0)/M_{\rm AFM}(T)$.
This includes a correction term to correspond to the case of the fully grown order parameter, because the data shown in Fig.~4(d) include those slightly below $T_{\rm N}$.
The $\sigma_{\rm AH}$ and corrected $\widetilde{M}  \sigma_{\rm AH}$ show the obvious $\sigma$ dependence, indicating that the extrinsic contribution was contained.
Previous systematical investigations of Fe films have clarified that the extrinsic contribution $\sigma_{\rm AH}^{ext}$, which includes both skew- and side-jump scattering, follows $\sigma_{\rm AH}^{ext} \propto \sigma^2$ in the temperature variation \cite{Tian}.
This corresponds to $\rho_{\rm AH}^{ext}$ being temperature independent. 
As the black curve shows in Fig.~4(d), $\widetilde{M} \sigma_{\rm AH}$ for Ce$_2$CuGe$_6$ followed the summation of two contributions, $\sigma_{\rm AH}^{int}+\sigma_{\rm AH}^{ext}$.
The $\sigma_{\rm AH}^{int}$ was estimated to be $\sim80$ $\Omega^{-1}$cm$^{-1}$ (dotted red line), whereas $\sigma_{\rm AH}^{ext}$ was $5\times10^{-9} \sigma^2$ (solid red line).
This extrinsic term corresponds to $\rho_{\rm AH}^{ext}=-5\times10^{-3}$ $\mu \Omega$cm, which should be temperature-independent at low temperatures.
This analysis suggests that $\rho_{\rm AH}$ shown in Fig.~4(b) was dominated by the extrinsic contribution at low temperatures. 
The intrinsic contribution in $\rho_{\rm AH}$ is significant just below $T_{\rm N}$ and is suppressed at low temperatures because of $\rho^{int}_{\rm AH} \propto \rho^2$. 
In Fig.~4(c), the intrinsic and extrinsic contributions in $\sigma_{\rm AH}$, which are estimated under the assumption of $\sigma_{\rm AH} \propto M_{\rm AFM}(T) \sigma^n$, are shown.
The extrinsic contribution appears at low temperatures, being consistent with the two-step development of $\sigma_{\rm AH}$.

Both the extrinsic and intrinsic mechanisms deriving from the AFM structure are allowed when their symmetries are equivalent to those of ferromagnets \cite{Chen14}.
A study in the Fe films has suggested that $\rho_{\rm AH}^{ext}$ is determined by the residual resistivity $\rho_0$ \cite{Tian}.
For Ce$_2$CuGe$_6$, $\rho_{\rm AH}^{ext}=-5\times10^{-3}$ $\mu \Omega$cm was obtained for $\rho_0\sim 3$ $\mu \Omega$cm.
For the Fe film, $\rho_0= 3$ $\mu \Omega$cm yielded $\rho_{\rm AH}^{ext}=-4.8\times 10^{-3}$ $\mu \Omega$cm, which consists of skew-scattering contribution $\rho_{\rm AH}^{skew}=-11.1 \times 10^{-3}$ $\mu \Omega$cm and side-jump contribution $\rho_{\rm AH}^{side}=6.3 \times 10^{-3}$ $\mu \Omega$cm \cite{Tian}.
This comparison means that the extrinsic contribution in Ce$_2$CuGe$_6$ is comparable to that of the conventional itinerant ferromagnet, although the net magnetization is two orders smaller.
The absence of the obvious AHE in $T=$ Au clearly shows that the AHE in Ce$_2$CuGe$_6$ is not relevant to the net magnetization.
Therefore, the observed AHE was considered as deriving from the AFM structure through both intrinsic and extrinsic mechanisms.
This study suggests that the conduction bands in these systems show spin splitting, which reflects the symmetry of the AFM structure, via the exchange interaction with the $f$ electrons. The opposite sign in $\rho_{\rm AH}$ between $T=$ Cu and Pd indicates a difference in the band topology.  Investigations of the band structure and its topology are crucial to understanding the mechanism of the observed AHE.

In conclusion, we investigated AHE for Ce$_2$CuGe$_6$, whose AFM structure is described by the FM magnetic point group, together with isostructural Ce$_2$PdGe$_6$ and Ce$_2$AuGe$_6$.
We observed the clear zero-field AHE accompanied by hysteresis loops for Ce$_2$CuGe$_6$ and Ce$_2$PdGe$_6$. 
Despite the weak net magnetization, the estimated AHC of $\sim550$ $\Omega^{-1}$cm$^{-1}$ was comparable to those of itinerant ferromagnets.
Careful analysis suggested that both the intrinsic and extrinsic mechanisms contributed to the AHE in Ce$_2$CuGe$_6$, where the intrinsic AHC was estimated to be $\sim80$ $\Omega^{-1}$cm$^{-1}$.
The following two important findings were derived from this study.
First, the relatively large intrinsic contribution emerged even from the localized $f$-electron AFM structure.
Second, the extrinsic contribution from the AFM structure was confirmed to be comparable to that observed in ferromagnets.
These findings can act as guidelines for obtaining a fuller understanding of the AHE and motivate further development of functional materials that generate FM-like responses in $f$-electron systems.

\section*{Acknowledgments}
We thank Hisatomo Harima, Michi-To Suzuki, and Shingo Araki for valuable discussions and comments. 
This work was supported by JSPS KAKENHI Grant Nos. 21K03446 and 23H04871 and by the Murata Science Foundation.

\end{document}